\documentclass[preprint]{aastex631}
\usepackage{bm}
\usepackage{comment}
\usepackage{color}
\usepackage{url}
\submitjournal{ApJ}

\shorttitle{Data-constrained MHD simulation of X1.0 flare}
\shortauthors{Yamasaki et al.}

\begin{document}

\title{A Data-constrained Magnetohydrodynamic Simulation of the X1.0 Solar Flare of 2021 October 28}

\correspondingauthor{Daiki Yamasaki}
\email{dyamasaki@kusastro.kyoto-u.ac.jp}

\author{Daiki Yamasaki}
\affiliation{Astronomical Observatory, Kyoto University \\
Kitashirakawaoiwake-cho, Sakyo-ku, Kyoto, 606-8502, Japan}
\affiliation{Center for Solar-Terrestrial Research, New Jersey Institute of Technology, \\
  University Heights, Newark, NJ 07102-1938, USA}

\author{Satoshi Inoue}
\affiliation{Center for Solar-Terrestrial Research, New Jersey Institute of Technology, \\
  University Heights, Newark, NJ 07102-1938, USA}

\author{Yumi Bamba}
\affiliation{Institute for Advanced Research (IAR), Nagoya University, \\
  Furo-cho, Chikusa-ku, Nagoya 464-8601, Japan}
\affiliation{Institute for Space-Earth Environmental Research (ISEE), Nagoya University, \\
  Furo-cho, Chikusa-ku, Nagoya 464-8601, Japan}

\author{Jeongwoo Lee}
\affiliation{Center for Solar-Terrestrial Research, New Jersey Institute of Technology, \\
  University Heights, Newark, NJ 07102-1938, USA}

\author{Haimin Wang}
\affiliation{Center for Solar-Terrestrial Research, New Jersey Institute of Technology, \\
  University Heights, Newark, NJ 07102-1938, USA}



\begin{abstract}

The solar active region NOAA 12887 produced a strong X1.0 flare on 2021 October 28, which exhibits X-shaped flare ribbons and a circle-shaped erupting filament. To understand the eruption process with these characteristics, we conducted a data constrained  magnetohydrodynamics simulation using a nonlinear force-free field of the active region about an hour before the flare as the initial condition. Our simulation reproduces the filament eruption observed in the H$\alpha$ images of GONG and the 304 $\mathrm{\AA}$ images of SDO/AIA and suggests that two mechanisms can possibly contribute to the magnetic eruption. One is the torus instability of the pre-existing magnetic flux rope (MFR), and the other is upward pushing by magnetic loops newly formed below the MFR via continuous magnetic reconnection between two sheared magnetic arcades. The presence of this reconnection is evidenced by the SDO/AIA observations of the 1600 {\AA} brightening in the footpoints of the sheared arcades at the flare onset. To clarify which process is more essential for the eruption, we performed an experimental simulation in which the reconnection between the sheared field lines is suppressed. In this case too, the MFR could erupt, but at a much reduced rising speed. 
We interpret this result as indicating that the eruption is not only driven by the torus instability, 
but additionally accelerated by newly formed and rising magnetic loops under continuous reconnection.

\end{abstract}



\section{Introduction} \label{sec:intr}
Solar flares are the rapid energy release phenomena in solar atmosphere \citep{Priest2002}.
The energy source of solar flares is widely considered as magnetic energy accumlated in solar active regions \citep[ARs;][]{Toriumi2019}.
Some of the free magnetic energy are converted into kinetic energy of erupting plasmas through magnetic reconnection \citep{Coppi1971,Spicer1982,Shibata2011}. 
These erupting plasmas are often observed as filament eruptions in the H$\alpha$ line \citep{Parenti2014,Seki2017,Seki2019}.
Solar filaments are cool and dense plasma compared to the surrounding plasma, they are widely considered to be supported by helical coronal magnetic field structure, magnetic flux ropes \citep[MFRs;][]{Xu2012,Hanaoka2017,Gibson2018}.

\citet{Chintzoglou2019} proposed that emerging dipole flux become strongly sheared due to photospheric motion and result in the formation of a bundle of helical magnetic field lines.
\citet{Yan2016} presented an observation that magnetic reconnection between pre-existing sheared magnetic arcades forms MFRs.
This process is consistent with the tether-cutting reconnection scenario by \citet{Moore2001} and the flux-cancellation model by \citet{vanBallegooijen1989}.
In solar eruptions, several triggering processes are proposed, for instance, the MHD instability such as the kink instability \citep{Fan2003,Toeroek2004} and double-arc instability \citep{Ishiguro2017,Kusano2020}, or magnetic reconnection \citep{Moore2001,Antiochos1999}.
To accelerate the solar eruption, the torus instability \citep{Kliem2006} plays an important role.

In order to understand the initiation of MFR eruptions, three-dimensional (3D) coronal magnetic field provides crucial information because the free magnetic energy is released in the solar corona and topology of the coronal magnetic field changes associated with a flare.
Observational limitation cannot allow us for direct observation of coronal magnetic field, and we need to extrapolate them by using a numerical technique such as potential field and nonlinear force-free field (NLFFF) modeling \citep{Wiegelmann2012,Inoue2016}.
There are several studies using a time series of NLFFF to reveal the formation process of MFRs \citep{SuY2009,Savcheva2009,Inoue2013,Kang2016,Kawabata2017,Muhamad2018,Yamasaki2021}. 
However, NLFFF is assumed as an equilibirum state only considering Lorentz force balance, and we cannot discuss dynamical evolutions of 3D magnetic field during eruptive phase of flares.
Since a data-constrained magnetohydrodynamic (MHD) simulation show time evolving MHD processes of the coronal magnetic field that are free from a force-free condition, we can extend our understanding to the dynamics of an erupting precess.
Some of the previous studies on the data-constrained MHD simulations successfully produced the eruptions that are driven by a reconnection or instability and these results are in good agreement with observations \citep{Jiang2013,Inoue2015,Inoue2018b,HeW2020,Inoue2021}.
Therefore, the data-based MHD simulation is helpful to understand dynamics of the coronal magnetic field in a realistic magnetic environment.

In this study, in order to understand the initiation and dynamics of the MFR eruption associated with X1.0 flare on October 28 2021, we performed a data-constrained MHD simulation using a NLFFF as the initial condition.
The NLFFF is extrapolated with observed photospheric vector magnetic field at 14:00 UT October 28, which is 1.5 hours before the onset of the flare.
The rest of this paper is structured as follows:
the observations and methods of analysis are described in Section \ref{sec:meth}, results are presented in Section \ref{sec:resu}, discussions on the eruption mechanism of the MFR are shown in Section \ref{sec:disc}, and our conclusions are summarised in Section \ref{sec:conc}.

\begin{figure*}[htb]
  \begin{center}
    \includegraphics[bb= 0 0 1225 770, width=150mm]{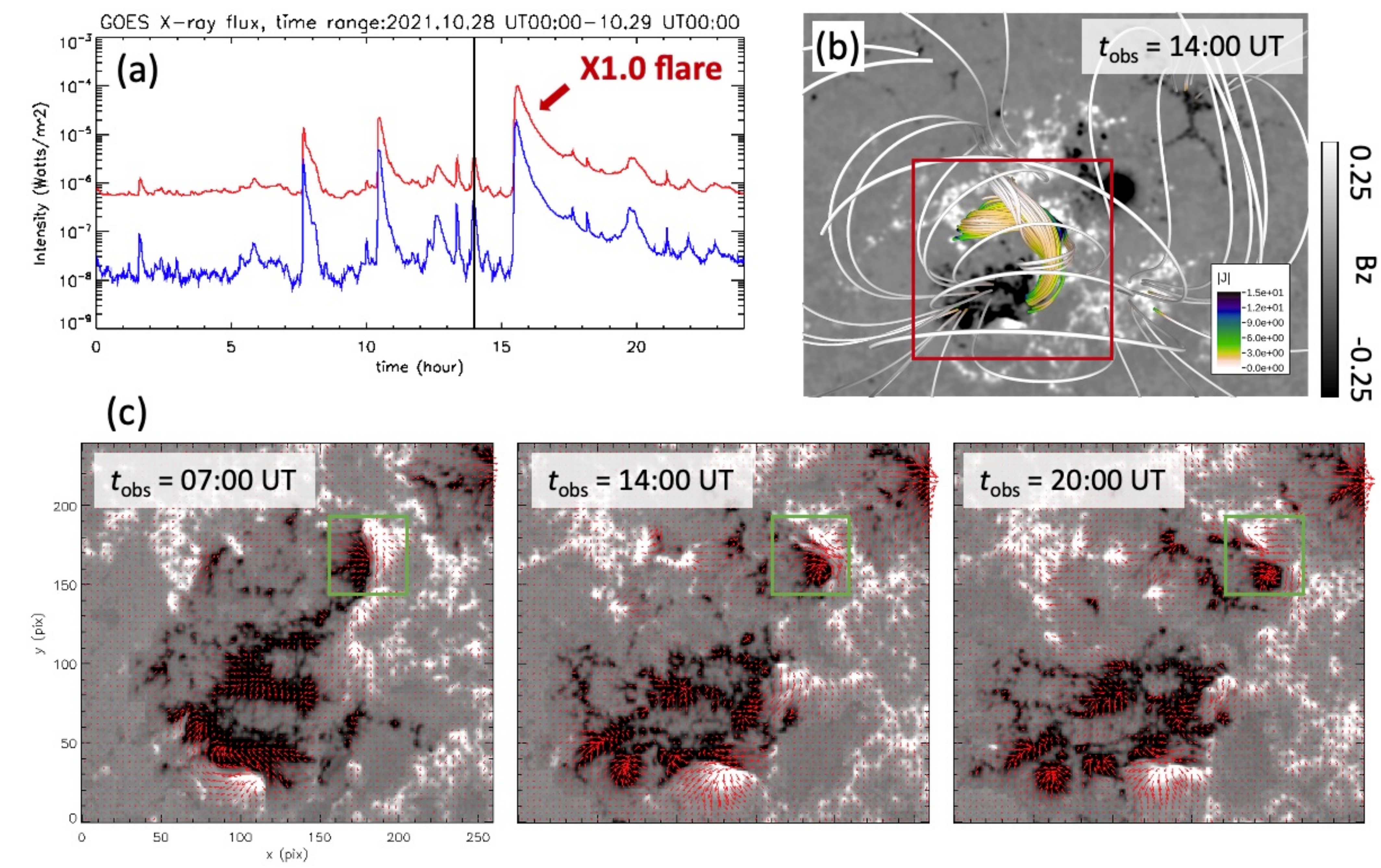}
    \caption{The 2021 October 28 flare in GOES soft X-ray lightcurves and HMI vector magnetograms. 
      (a) Soft X-ray fluxes from the $GOES$ 13 satellite in the $1-8\mathrm{\AA}$ (red) and $0.5-4.0\mathrm{\AA}$ (blue) passband. The vertical black line shows the time of the initial condition set for the present simulation. (b) Radial component of the photospheric magnetic field in the AR 12887 at 14:00 UT on October 28. Color of the field lines represent the electric current density. (c) Temporal evolution of the photospheric vector magnetic field. Gray-scaled background and red arrows show the radial and horizontal component of the magnetic field, respectively.
 (An animation of the HMI photospheric magnetic field images is available. The duration of the animation is 4 seconds and it provides information of the temporal evolution of the photospheric vector magnetic field including an intruding motion of a negative patch into a positive patch from 00:00 UT to 23:48 UT 2021 October 28.)}
    \label{fig1}
  \end{center}
\end{figure*}

\section{Observation} \label{subsec:obse}
The GOES class X1.0 flare occurred on 2021 October 28 in NOAA active region (AR) 12887. According to the $GOES$ X-ray lightcurves in Figure \ref{fig1} (a), the onset time was 15:17 UT and the soft X-ray flux reached its maximum at 15:35 UT. The black solid line at 14:00 UT indicates the time that we selected for calculating the bottom boundary to be used for the MHD simulation. 
Figure \ref{fig1} (b) shows the radial magnetogram and the coronal magnetic field lines extrapolated with NLFFF around the AR 12887 at 14:00 UT.
The photospheric vector magnetic field data are taken by the Helioseismic and Magnetic Imager \citep[HMI; ][]{Scherrer2012} onboard the $Solar~ Dynamics~ Observatory$ \citep[$SDO$; ][]{Pesnell2012}. 
The vector magnetograms which we use have been released as the Spaceweather HMI Active Region Patch \citep[SHARP; ][]{Bobra2014} data series (hmi.sharp\_cea\_720s series).
Details of the vector magnetic field data reduction and other related information about HMI data products can be found in \citet{Hoeksema2014} and \citet{Bobra2014}.
In Figure \ref{fig1} (c), we show the temporal evolution of the photospheric vector magnetic field in the region of interest, which is the same as the region indicated by the red box in Figure \ref{fig1} (b).
Background grayscale and the red arrows show the radial and the horizontal component of the magnetic field, respectively.
According to Figure \ref{fig1} (c), we can find an intruding motion of the negative polarity towards the positive polarity on October 28 in the region indicated by the green box in the panels. 

\begin{figure*}[htb]
  \begin{center}
    \includegraphics[bb= 0 0 1430 990, width=150mm]{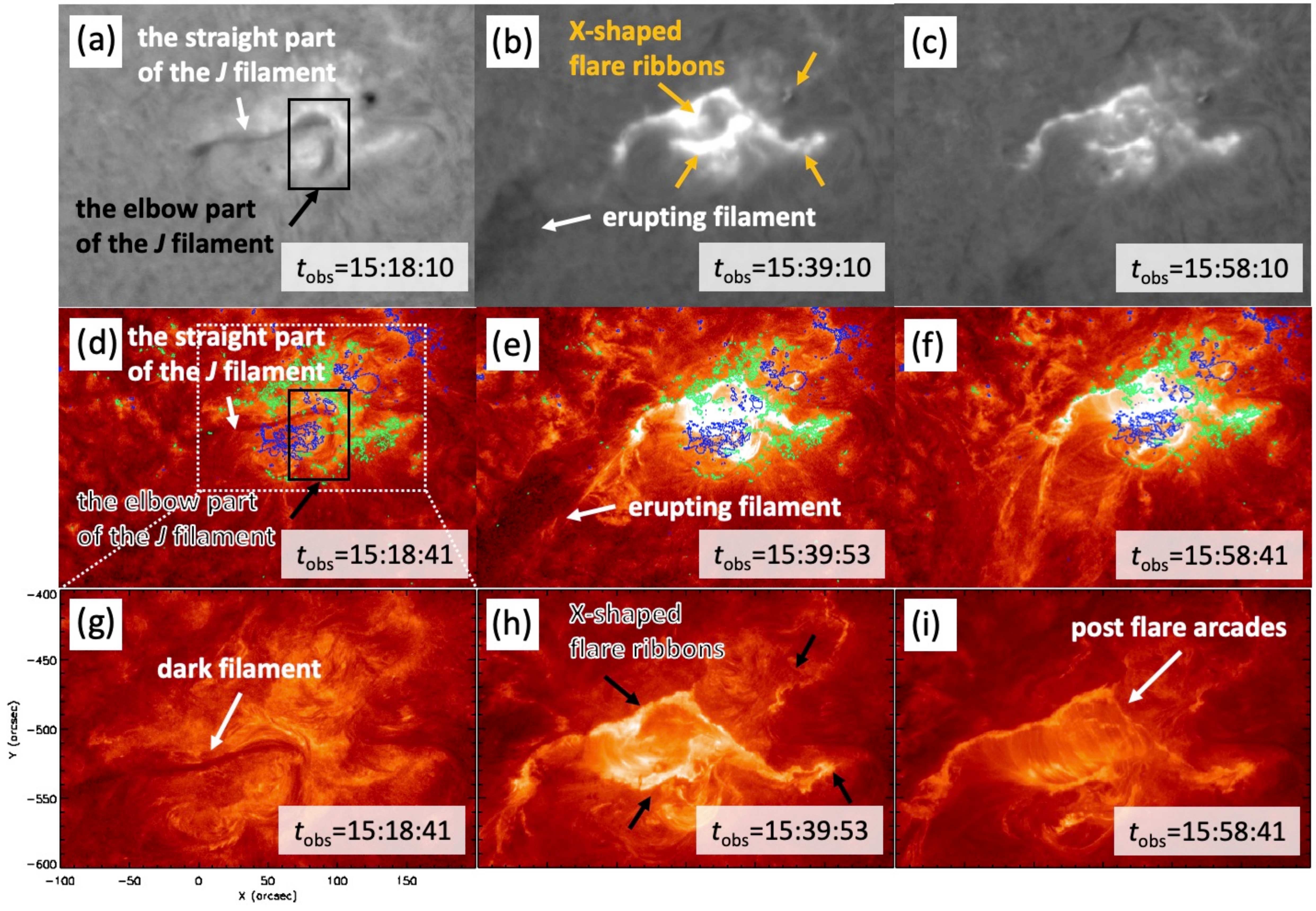}
    \caption{Temporal evolution of AR 12887 during the 2021 October 28 flare. (a-c) H$\alpha$ (6562.8 $\mathrm{\AA}$) observation with GONG shows the flare ribbons and a dark filament that later erupts. (d-i) SDO/AIA 304 $\mathrm{\AA}$ images show the flare loops, the erupting filament, and post-flare arcades. Green and blue lines in panels (d-f) indicate the contours of 250 and -250 G of radial component of the photospheric magnetic fields, respectively.
    (An animation of the AIA 304 images is available. The duration of the animation is 5 seconds and it provides information on the temporal evolution of dark filament eruption, flare enhacements, and post flare arcades from 15:00 UT to 16:00 UT 2021 October 28.)}
    \label{fig2}
  \end{center}
\end{figure*}

Figure \ref{fig2} (a-c) show the temporal evolution of the H$\alpha$ (6562.8 $\mathrm{\AA}$) images observed with the Global Oscillation Network Group \citep[GONG; ][]{Harvey1996} around the AR 12887 during the X1.0 flare.
Details of the instrument can be found in \citet{Harvey1995}.
Figure \ref{fig2} (d-i) show the temporal evolution of the 304 $\mathrm{\AA}$ images observed by the Atmospheric Imaging Assembly \citep[AIA; ][]{Lemen2012} on board the $SDO$ during the X1.0 flare.
Green and blue lines in panels (d-f) indicate the contours of 250 and -250 G of radial component of the photospheric magnetic fields observed by HMI, respectively.
We can find the J-shaped dark filament in the early phase of the flare (Figure \ref{fig2} (a,d,g)) and the erupting filament and the X-shaped flare ribbons in the main flare phase (Figure \ref{fig2} (b,e,h)).
The H$\alpha$ flare ribbons last until the later phase of the flare (Figure \ref{fig2} (c)).
The post flare arcades can be seen in the 304 $\mathrm{\AA}$ images (Figure \ref{fig2} (f,i)).

\section{Methods} \label{sec:meth}

We construct a nonlinear force-free field and then use it as the initial condition for the data-constrained MHD simulation. 
Once the 3D magnetic field is obtained, we calculate the magnetic twist number and the decay index to study the instability leading to eruption.

\subsection{Nonlinear Force-Free Field Extrapolation} \label{subsec:nlff}
In order to perform the NLFFF extrapolation, we first calculate the potential field \citep{sakurai1982} which is used as the initial condition of the MHD relaxation \citep{Inoue2014,Inoue2016}.
We solve zero-beta MHD equations to obtain the force-free field because the gas pressure and gravity are neglected approximately compare to the magnetic pressure in the  solar corona \citep{Gary2001}.
We solve the following MHD equations, 
\begin{eqnarray}
  \rho &=& |\bm{B}|, \label{eq1}\\
  \frac{\partial \bm{v}}{\partial t} &=& -(\bm{v}\cdot{\bm{\nabla}})\bm{v}+\frac{1}{\rho}\bm{J}\times\bm{B}+\nu_{i}\bm{\nabla}^{2}\bm{v}, \label{eq2}\\
  \frac{\partial \bm{B}}{\partial t} &=& \bm{\nabla}\times(\bm{v}\times\bm{B}-\eta_{i}\bm{J})-\bm{\nabla}\phi, \label{eq3}\\
  \bm{J} &=& \bm{\nabla}\times\bm{B},\\
  \frac{\partial \phi}{\partial t}&+&c_{\mathrm{h}}^{2}\bm{\nabla}\cdot\bm{B} = -\frac{c_{\mathrm{h}}^{2}}{c_{\mathrm{p}}^{2}}\phi, \label{eq5}
\end{eqnarray}
where the subscript $i$ of $\nu$ and $\eta$ corresponds to different values used in NLFFF and MHD.
$\rho$, $\bm{B}$, $\bm{v}$, $\bm{J}$, and $\phi$ are plasma pesudo-density, magnetic flux density, velocity, electric current density, and a conventional potential to reduce errors derived from $\bm{\nabla}\cdot\bm{B}$ \citep{Dedner2002}. 
The pseudo-density in equation (\ref{eq1}) is assumed to be proportional to $|\bm{B}|$.
In these equations, the length, magnetic field, density, velocity, time, and electric current density are normalized by $L^*=254.8$ $\mathrm{Mm}$, $B^*=3000$ $\mathrm{G}$, $\rho^{*}=|B^{*}|$, $V_{\mathrm{A}}^{*}\equiv B^*/(\mu_{0}\rho^*)^{1/2}$, where $\mu_{0}$ is the magnetic permeability, $\tau_{\mathrm{A}}^{*}\equiv L^{*}/V_{\mathrm{A}}^{*}$, and $J^{*}=B^{*}/\mu_{0}L^{*}$, respectively.
In equation (\ref{eq2}), $\nu_{\mathrm{NLFFF}}$ is a viscosity fixed at $1.0\times10^3$.
The coefficients $c_{\mathrm{h}}^{2}, c_{\mathrm{p}}^{2}$ in equation (\ref{eq5}) are fixed to the constant values $0.04$ and $0.1$.
The resistivity in equation (\ref{eq3}) is given as $\eta_{\mathrm{NLFFF}}=\eta_{0}+\eta_{1}|\bm{J}\times\bm{B}||\bm{v}|^{2}/|\bm{B}|^{2}$, where $\eta_{0}=5.0\times10^{-5}$ and $\eta_{1}=1.0\times10^{-3}$ in non-dimensional units.
As for the boundary conditions, three components of the magnetic field are fixed at each boundary, while the velocity is fixed to zero and the von Neumann condition $\partial/\partial n=0$ is imposed on $\phi$ during the iteration. 
Here we note that we fixed the bottom boundary according to
\begin{eqnarray}
  \bm{B_{\mathrm{bc}}}=\gamma\bm{B_{\mathrm{obs}}}+(1-\gamma)\bm{B_{\mathrm{pot}}},
\end{eqnarray}
where $\bm{B_{\mathrm{bc}}}$ is the transversal component determined by a linear combination of the observed magnetic field ($\bm{B}_{\mathrm{obs}}$) and the potential magnetic field ($\bm{B_{\mathrm{pot}}}$).
$\gamma$ is a coefficient in the range of 0 to 1.
The value of the parameter $\gamma$ is increased to $\gamma=\gamma+d\gamma$ if $R=\int|\bm{J}\times\bm{B}|^{2}dV$, which is integrated over the computational domain, becomes smaller than a critical value which is denoted by $R_{\mathrm{min}}$ during the iteration. In this paper, we set $R_{\mathrm{min}}$ and $d\gamma$ the values of $5.0\times10^{-3}$ and $0.02$, respectively.
When $\gamma$ reaches to 1, $\bm{B_{\mathrm{bc}}}$ is completely consistent with the observed data.
Furthermore, we controll the velocity as follows. If the value of $v^{*}(=|\bm{v}|/|\bm{v_{A}}|)$ is larger than $v_{\mathrm{max}}$ (here we set to $0.04$), then we modify the velocity from $\bm{v}$ to $(v_{\mathrm{max}}/v^{*})\bm{v}$.
We adopted these processes because they would help avoid a sudden jump from the boundary into the domain during the iterations.

\subsection{Data-constrained MHD Simulation} \label{subsec:mhd}
Next we performed the MHD simulation using NLFFF as an initial condition.
We name this simulation RUN A. 
The equations we solve are identical to those in the NLFFF extrapolation.
However, the handling of the bottom boundary condition is different between the NLFFF extrapolation and the data-constrained simulation.
In the data-constrained simulation, the bottom $B_{\mathrm{x}}$ and $B_{\mathrm{y}}$ follow an induction equation while the normal  component is fixed with time.
Although the boundary magnetic field evolve with time in a physically consistent manner, these evolutions are inconsistent with the observations \citep{Inoue2021}.
We set resistivity and viscosity as $\eta_{\mathrm{MHD}}=1.0\times10^{-4}$ and $\nu_{\mathrm{MHD}}=1.0\times10^{-3}$, respectively, which are different from those in the NLFFF.
The coefficients $c_{\mathrm{h}}^{2}$ and $c_{\mathrm{p}}^{2}$ in equation (\ref{eq5}) are fixed to the constant values $0.04$ and $0.1$, respectively.

For both calculations, the numerical domain has dimensions of $255\times195\times191$ $\mathrm{Mm}^{3}$, or $1.00\times0.77\times0.75$ in non-dimensional units.
The region is devided into $352\times270\times264$ grid points. 

\subsection{Analysis of Magnetic Fields} 
Once the 3D magnetic field is obtained, we calculate the magnetic twist number and the decay index to study the instability leading to eruption.
The magnetic twist of each field line \citep[$e.g.$ ][ etc.]{Inoue2011,Liu2016} is calculated using the following definition \citep{Berger2006},
\begin{eqnarray}
  T_{\mathrm{w}} = \int_{L}\frac{\bm{\nabla}\times\bm{B}\cdot\bm{B}}{4\pi B^2}~ dl,  
\end{eqnarray}
where $dl$ is a line element. 
The decay index, $n$, the proxy criterion for the torus instability \citep{Kliem2006} is calculated as
\begin{eqnarray}
  n=-\frac{z}{|\bm{B_{\mathrm{ex}}}|}\frac{\partial |\bm{B_{\mathrm{ex}}}|}{\partial z}
\end{eqnarray}
Here $\bm{B_{\mathrm{ex}}}$ denotes the horizontal component of the external field, which is assumed to be the potential field in this study. 

\section{Results} \label{sec:resu}

\subsection{Evolution of the MFR}
\begin{figure*}[htb]
  \begin{center}
    \includegraphics[bb= 0 0 1270 685, width=150mm]{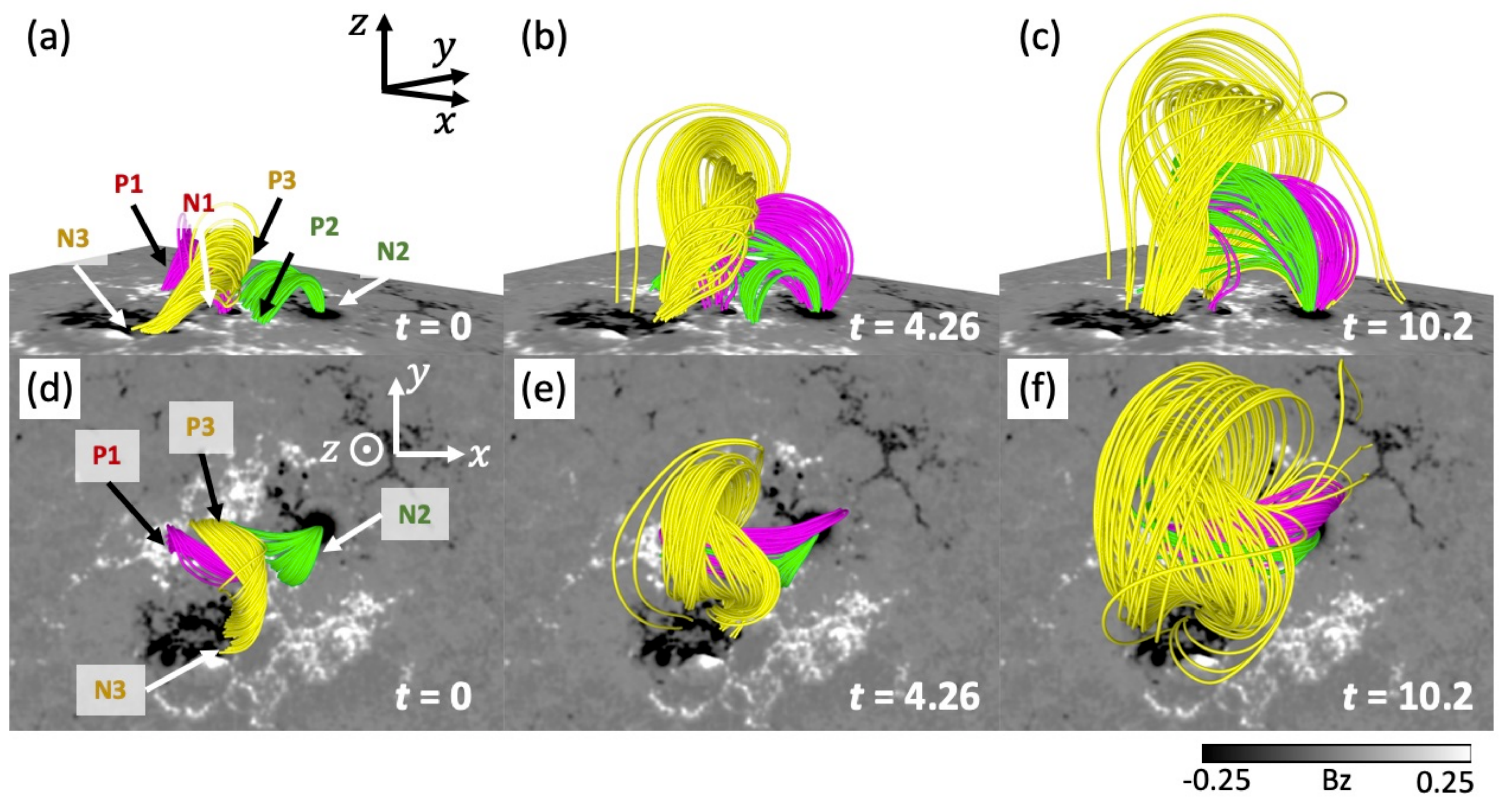}
    \caption{Three snapshots of the simulation of RUN A, showing temporal evolution of the coronal magnetic fields. (a) and (d) show top and side views of the 3D magnetic fields at $t=0$. (b) and (e) show those at $t=4.26$. (c) and (f) show those at $t=10.2$.  The pink, green, and yellow lines are the field lines with footpoints at P1-N1, P2-N2, and P3-N3, respectively. The yellow lines are identified as an MFR that erupts. 
    (An animation of the RUN A simulation is available. The duration of the animation is 1 second and it provides information on the temporal evolution of the 3D magnetic field in our RUN A simulation including the MFR eruption and magnetic reconnection between pre-existing sheared magnetic arcades.)}\label{fig3}
  \end{center}
\end{figure*}

In Figure \ref{fig3}, we display three snapshots of the MHD simulation to show temporal evolution of the coronal magnetic field.
Figure \ref{fig3} (a) and (d) show the initial condition of the simulation viewed in two different perspective angles.
Figure \ref{fig3} (b,e) and (c,f) show the magnetic field structure at $4.26$ and $t=10.2$ at the same angle as above, respectively.
The colored field lines are selected in the criteria of strong twist built up during the evolution.
For this purpose, we first calculated the magnetic twist of every field line starting from each pixel using equation (7), which yields a map of magnetic twist on the bottom boundary.
We then select the regions with $T_{\mathrm{w}}>1.0$ at the final time step ($t=10.2$) of the simulation from which we perform the field line tracing.
P1, P2, and P3 correspond to the footpoints of the pink, green, and yellow field lines in the regions of positive magnetic polarity, and N1, N2, N3 are their counterparts in the negative polarity.
In our NLFFF, only the elbow part of the J-shaped dark filament located at the west side of the AR (Figure \ref{fig2} (a)) could be well reconstructed by the yellow field lines in Figure \ref{fig3}. 
We think that this is because the straight part of the J-shaped filament formed at the east side of the AR was located above the relatively weak field region and failed to build a highly twisted structure.
We define a bundle of field lines with $T_{\mathrm{w}}>1.0$ as an MFR and that with $0<T_{\mathrm{w}}<1.0$ as a sheared magnetic arcade. In this definition, the yellow lines are identified with an MFR, and pink and green lines, a pair of sheared magnetic arcades at the initial time.
In Figure \ref{fig3} (c,f), we can see the MFR (yellow lines) undergo eruption and expansion.
We suggest that the continuous magnetic reconnection taking place between the pink and green field lines led to the formation of magnetic loops below the MFR, see the online animation.

\subsection{Decay index distribution} \label{subsec:fluc}
\begin{figure*}[htb]
  \begin{center}
    \includegraphics[bb= 0 0 860 705, width=150mm]{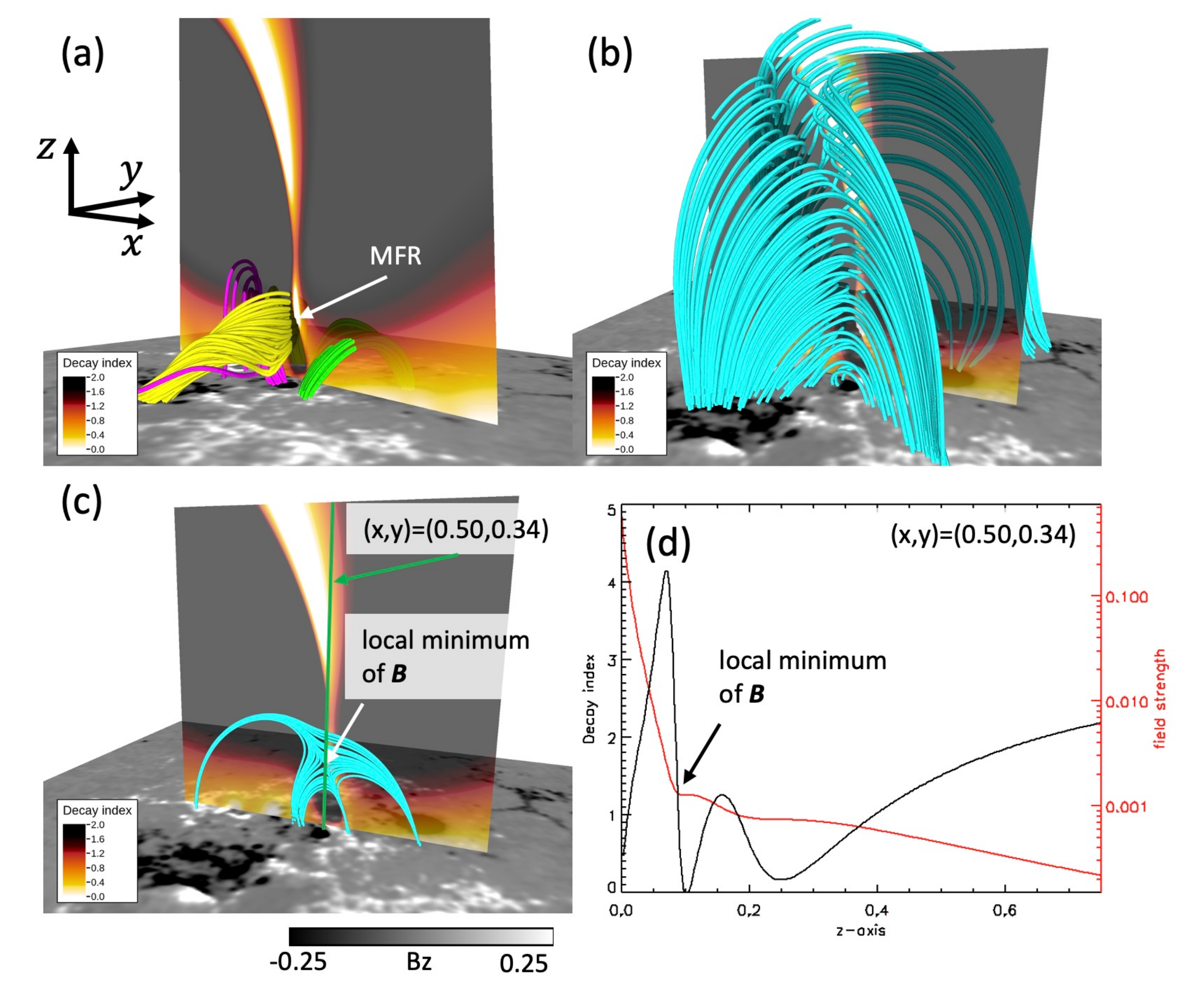}
    \caption{Decay index distribution in 3D. (a) The index is plotted in vertical cross section of $x$--$z$ plane at $y=0.34$. The initial magnetic field lines ($t=0$) in the simulation are plotted with the same color convention as used in Figure 3 (b). (b) Azure lines are the potential field lines surrounding the MFR. (c) The field lines of the potential field are projected on the $x$--$z$ plane at $y=0.34$. The green line is the vertical line of $(x,y)=(0.50,0.34)$ selected for plotting 1D distribution of the decay index and the magnetic field strength in (d). }\label{fig4}
  \end{center}
\end{figure*}

In Figure \ref{fig4} we plot the height distribution of the decay index calculated using equation (8) on a vertical $x-z$ plane at $y=0.34$.
Figure \ref{fig4} (a) shows that the MFR (yellow lines) was already located in the region of high decay index $n>2.0$ at the start.
In panels (b) and (c), we plot, in azure color, the potential magnetic field lines surrounding the MFR in 3D and 2D, respectively.
A magnetic null like region ($B\approx 0$) is found on the plane at a low latitude $z=0.09$ (Figure \ref{fig4} (c,d)).
Since magnetic field rapidly decreases toward the null, a large value of the decay index was realized in the low latitude (Figure \ref{fig4} (d)).
We, therefore, expect that the MFR (yellow lines) could easily become unstable to the torus instability.
In addition, as shown in panel (d), we find a torus stable region with $n<1.3$ above the initial location of the MFR ($0.09<z<0.46$). 
\citet{Kliem2021} pointed out that in the case of such cubic functional decay index distribution, an eruption starts but fails because of torus stable region. 
However, some numerical studies reported that the MFRs could erupt if they were located at the torus unstable region in the initial condition \citep{Inoue2018a,Zhong2021,Joshi2021}. 
In such a case, even if the decay index above the MFRs is low enough to be torus stable, the MFRs could erupt.
Our result supports the latter scenario. 

\subsection{Comparison of the flare ribbon structures of observation and simulation}
\begin{figure*}[htb]
  \begin{center}
    \includegraphics[bb= 0 0 930 910, width=150mm]{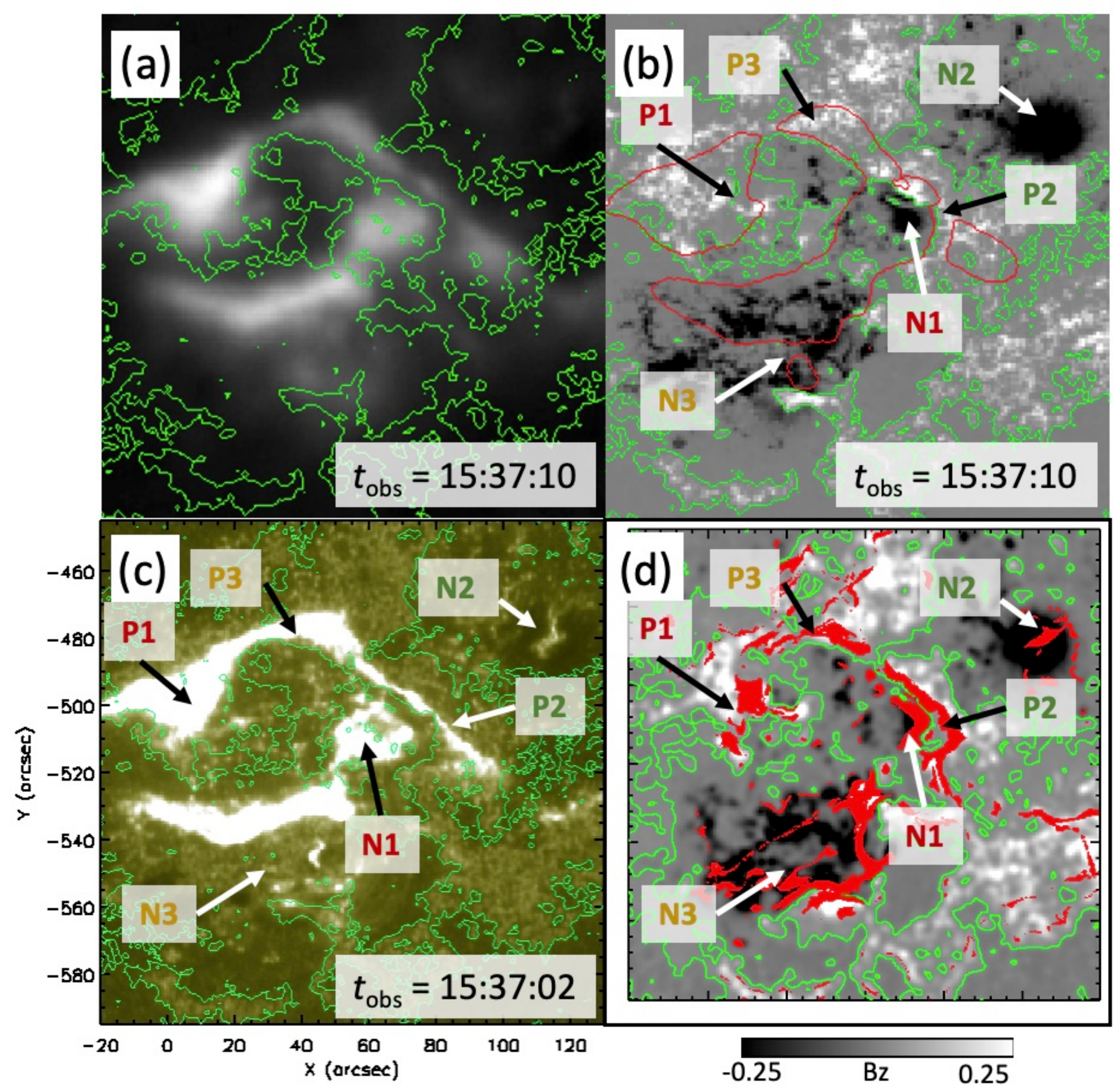}
    \caption{Observed and computed flare ribbons. Green lines in all the panels show the polarity inversion lines. (a) Flare ribbons in the GONG H$\alpha$ line image at 15:37:10 UT. (b) The radial component of the photospheric magnetic field overlaid with the enhanced H$\alpha$ intensity as red contours at 15:37:10 UT. (c) Flare ribbons in the AIA 1600 {\AA} channel at 15:37:02 UT. (d) The computed flare ribbons from the MHD simulation of RUN A (red) plotted over the magnetogram (grayscale).}\label{fig7}
  \end{center}
\end{figure*}

Figure \ref{fig7} shows the flare ribbons observed in H$\alpha$ and EUV (a-c) and calculated from the MHD model (d) in the main phase of the flare.
The H$\alpha$ and EUV images show bright ribbons not only at two footpoints of the MFR (P3 and N3), but also at other four footpoints of sheared magnetic arcades formed below the MFR (P1, P2, N1, and N2) (Figure \ref{fig7} (a-c)). 
The model flare ribbon (red features in Figure \ref{fig7} (d)) is calculated simply based on the distance between two footpoints of a field line per pixel. Since the only way that this length changes significantly is via reconnection, we regard those regions as ribbons. Specifically those pixels where field lines with the distance changing by more than $3.6$ $\mathrm{Mm}$ are marked red.
These computed flare ribbons are found not only at the footpoints of the MFR, but also at the four footpoints of sheared magnetic arcades consistent with the observation. This result supports the scenario of the reconnection between two sheared magnetic arcades below the pre-existing MFR during the X1.0 flare.

Although the computed flare ribbons in the northern part of the AR fairly well agree with the observed ribbons in both location and intensity, those in the southern part of the AR do not.
The observed ribbon in the location of $-$540$'' < y<-$530$''$ is barely predicted by the model, while the observed ribbon in $-$560$'' < y<-$550$''$ is very weak compared with the model prediction.
We consider two possibilities for the partial success of our ribbon prediction.
One is that our model prediction for flare ribbons is designed mainly for location but not intensity, and our criterion for ribbons based on the field line length change does not work well for the southern ribbons.
The other possibility is that the topology of the magnetic field in the southern area is not correctly reproduced under the NLFFF  extrapolation to limit the accuracy of our ribbon prediction in that area.

\section{Discussion} \label{sec:disc}
The results of simulation RUN A indicate two possibilities for the driving mechanism of the MFR eruption.
One is the torus instability of the MFR (Figure \ref{fig4}).
According to the result of the statistical study of an eruption mechanism done by \citet{Jing2018}, the decay index value of $n>2.0$ is large enough to produce eruptions via torus instability.
The other possibility is that the MFR is pushed upward by the magnetic loops formed by the magnetic reconnection between two sheared magnetic arcades below the pre-existing MFR (Figure \ref{fig3}).
This process can also drive the MFR eruption.
The newly formed loops are created through the reconnection, therefore it can be interpreted that the reconnection drives the MFR eruption.

\begin{figure*}[htb]
  \begin{center}
    \includegraphics[bb= 0 0 1360 505, width=150mm]{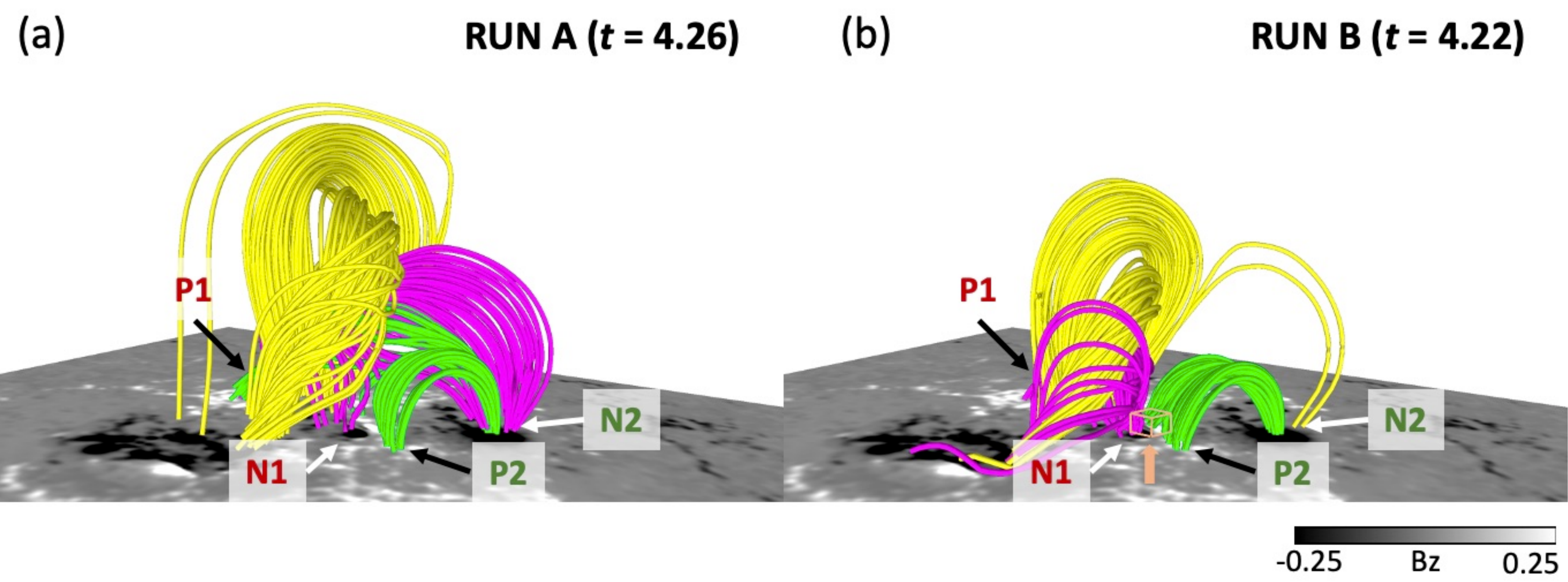}
    \caption{Comparison of magnetic fields in RUN A and RUN B. The magnetic fields at $t=4.26$ in RUN A (a) and those at $t=4.22$ in RUN B (b). The same color convention as in Figure 3 is used to identify the three groups of field lines. Orange box indicated by orange arrow in (b) shows the region where we set the velocity to zero in RUN B. It lies in $0.45\leq x\leq0.54$, $0.31\leq y\leq0.40$, and $0.02\leq z\leq0.06$, satisfying $|{\bm J}|\geq10$.}\label{fig5}
  \end{center}
\end{figure*}

In order to find out which process is essential for the eruption of the MFR, we conduct another MHD simulation named RUN B.
To clarify what role is played by the reconnection under the erupting filament, we halt the motion in the specific region where strong current density is formed; it lies between the sheared field lines in pink and green (see orange box in Figure \ref{fig5} (b)). 
This process is expected to partially suppress the magnetic reconnection between two sheared magnetic arcades. We thus prevent in this experiment, the large loops from  forming that is suspected to push the erupting MFR.

In Figure \ref{fig5}, we compare the field lines found in simulation RUN A with those in RUN B. The same color code is used for both.
The results of RUN B show how the field line structure differs when the reconnection between pink and green lines is suppressed.
However, the MFR with yellow lines moved upwards in RUN B as well as in RUN A (Figure \ref{fig5} (a,b)).
Therefore, the MFR can ascend without being pushed up by the large arcade.

\begin{figure*}[htb]
  \begin{center}
    \includegraphics[bb= 0 0 1160 560, width=150mm]{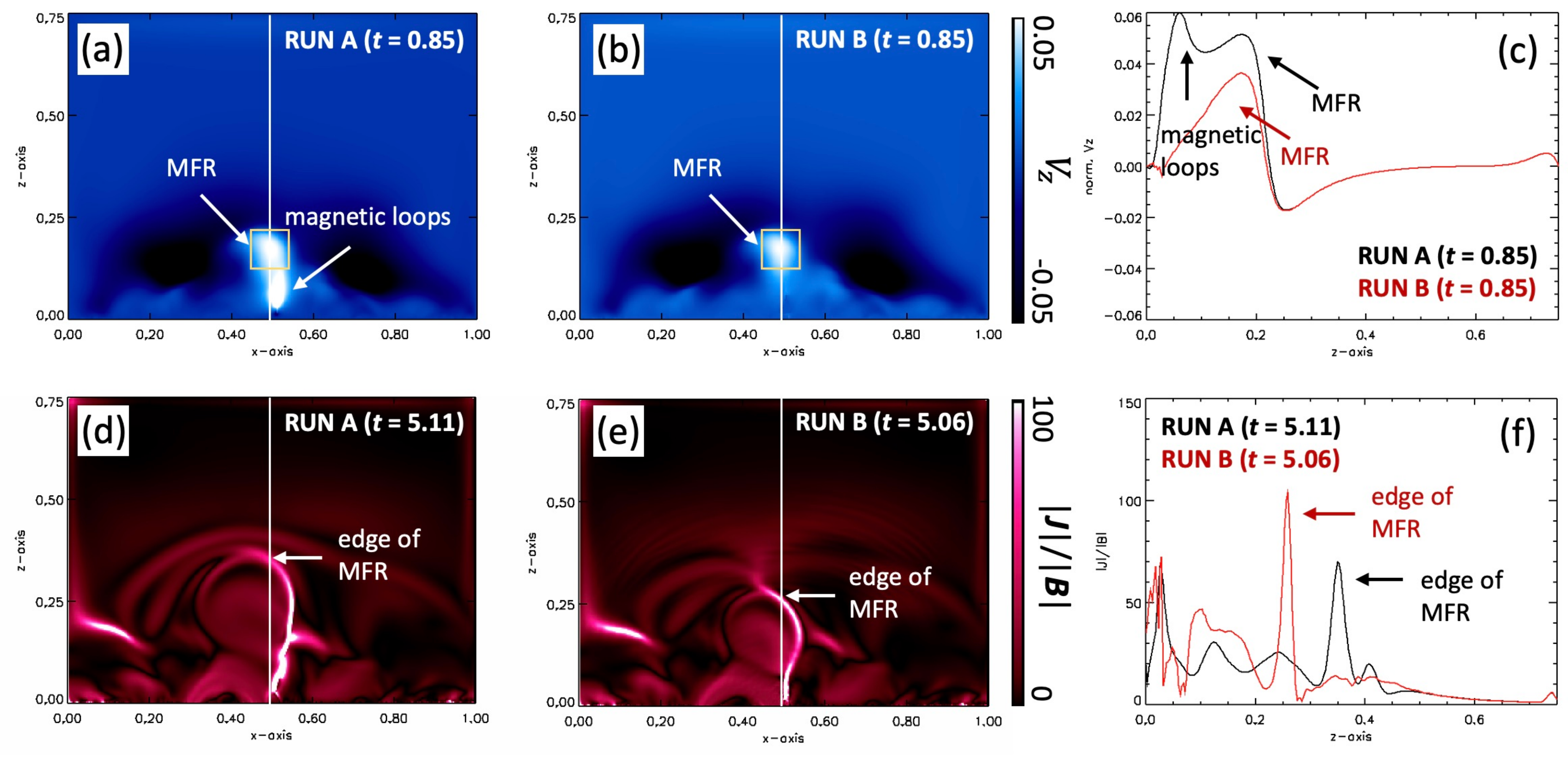}
    \caption{Comparison of RUN A and RUN B in terms of the vertical velocity, $V_z$ (a--c) and current density normalized by magnetic field, $|\bm{J}|/|\bm{B}|$ (d--f). (a,b) show 2D distribution of $V_z$ in a vertical $x-z$ plane at $y=0.34$, obtained from RUN A and RUN B at $t=0.85$, respectively. (c) shows their 1D distributions at the location of $(x,y)=(0.50,0.34)$ indicated by the white line marked in other panels. 
    Likewise, (d, e) show the 2D distribution, and (f), the 1D distribution of $|\bm{J}|/|\bm{B}|$ obtained from RUN A at $t=5.11$ and RUN B at $t=5.06$, respectively.
    Regions of enhanced $|\bm{J}|/|\bm{B}|$ are regarded as edges of the MFR.}\label{fig6}
  \end{center}
\end{figure*}

Figure  \ref{fig6} compares simulation RUN B with RUN A in terms of the vertical velocity ($V_{\mathrm{z}}$) and normalized current density, $|\bm{J}|/|\bm{B}|$.
The distribution of $V_{\mathrm{z}}$ in an $x$-$z$ plane is plotted at $y=0.34$ at a time in the early phase, $t=0.85$, for both simulations. 
RUN A shows that the enhanced upward $V_{\mathrm{z}}$ extends down to a lower height than in RUN B (Figure \ref{fig6} (a,b)).
In RUN A, the MFR and the newly formed magnetic loops exist in the region of enhanced velocity, whereas, in RUN B, the velocity is enhanced only in the location of the MFR.
The height-variation of $V_z$ plotted in Figure \ref{fig6} (c) clearly shows that the velocity is much more enhanced when the reconnection between the sheared field lines is allowed.

In a region of strongly enhanced $|\bm{J}|/|\bm{B}|$, the magnetic field topology may change rapidly and thus it could be a boundary between two topologically different regions. For instance, the region indicated by the white arrow in Figure \ref{fig6} (d,e) can indicate a boundary between the MFR and the overlying field lines, thus roughly the edge of the MFR.
In Figure \ref{fig6} (f), we show the one-dimensional plot of $|\bm{J}|/|\bm{B}|$.
The edge of the MFR in RUN A is higher than that in RUN B, while $V_{\mathrm{z}}$ of RUN A is still larger than that of RUN B.
These results suggest that the MFR could erupt even without being pushed by the magnetic loops newly formed below the MFR.
However, the push-up from below can be important to help the acceleration of the MFR.
And according to the results of the observed and the computed flare ribbons, they were observed not only at the two footpoints of the erupting MFR but also at the four footpoints of the reconnecting pre-existing magnetic arcades, the scenario of RUN A is more consistent with the observation rather than RUN B.

As shown in Figure \ref{fig5}, observation supports the scenario with magnetic reconnection between pink and green field lines. 
By comparing the photospheric magnetic field structure before and after the X1.0 flare, intruding motion of the negative polarity (N1 in Figure \ref{fig7}) into the positive polarity (P2 in Figure \ref{fig7}) was observed at the location where we found the magnetic reconnection in RUN A (see the green box of Figure \ref{fig1} (c)). 
\citet{Inoue2021} pointed out that the intruding motion can be a trigger of magnetic reconnection concerning to the MFR eruption. 
We suggest that the intruding motion we found in this case of X1.0 flare could be a trigger of the magnetic reconnection.

\section{Conclusion} \label{sec:conc}

We presented a data-constrained MHD simulation to understand the formation and erupting process of an MFR associated with the X1.0 flare that occurred on 2021 October 28, which is characterized by an X-shaped flare ribbon and a nearly circular J-shaped filament. Our simulation is meant to reproduce the observed MFR eruption with the shape of the filament and the ribbons as constraint. According to the initial condition obtained from a nonlinear force-free field extrapolation, the MFR was initially in a region of sufficient decay index. Therefore the MFR could erupt under the torus instability alone. 
However, we found that the magnetic reconnection between two sheared magnetic arcades took place under the pre-existing MFR during the erupting phase of the MFR, and paid attention to the possible role of this reconnection of the underlying loops in facilitating the eruption of the MFR. 

By performing an experimental simulation in which this reconnection is suppressed, we found that the MFR is still able to erupt but at a reduced speed. The MFR erupts faster when the large magnetic loops form underneath.
The reconnection of the magnetic loops formed below the MFR is thus essential for accelerating the MFR.
In addition, this reconnection cannot be overlooked because flare ribbons were observed not only at the footpoints of MFR but also at those of the newly formed magnetic loops during the filament eruption.
From these results, we propose that the initial driving mechanism of the filament eruption associated with this GOES X1.0 flare was facilitated by the combined action of the torus instability and the formation of the magnetic loops through the reconnection below the pre-existing MFR. A similar idea was presented by \cite{Inoue2018b}.

This event produced a strong CME which has been well studied in many papers \citep{Xu2022,Hou2022,Papaioannou2022,Li2022}.  It is thus of new interest how this MFR grows into the CME, although the present simulation domain is not large enough to fully trace the MFR in a long range. It is also worthwhile to investigate how the pre-erupting MFR is formed. 
We plan to address these issues in future with an extended time coverage.

\begin{acknowledgements}
$SDO$ is a mission of NASA’s Living With a Star Program.
This work utilizes data from the National Solar Observatory Integrated Synoptic Program, which is operated by the Association of Universities for Research in Astronomy, under a cooperative agreement with the National Science Foundation and with additional financial support from the National Oceanic and Atmospheric Administration, the National Aeronautics and Space Administration, and the United States Air Force.
The GONG network of instruments is hosted by the Big Bear Solar Observatory, High Altitude Observatory, Learmonth Solar Observatory, Udaipur Solar Observatory, Instituto de Astrofísica de Canarias, and Cerro Tololo Interamerican Observatory.
This work was supported by MEXT/JSPS KAKENHI Grant Number JP21J14036 and 21K20379.
This work was also supported by JSPS Overseas Challenge Program for Young Researchers.
This study is partially supported by National Science Foundation AGS-2145253,  AGS-1954737 and AST-2204384. JL acknowledges support by NASA grants, 80NSSC18K1705 and 80NSSC21K1671.
Visualization of magnetic field lines are produced by VAPOR (\url{www.vapor.ucar.edu}), a product of the Computational Information Systems Laboratory at the National Center for Atmospheric Research \citep{atmos10090488}.
\end{acknowledgements}

\bibliography{Yamasaki2022c}
\bibliographystyle{aasjournal}

\end{document}